\documentclass[conference]{IEEEtran}
\usepackage{amsmath}
\usepackage{latexsym}
\usepackage{graphicx}
\usepackage{bbding}
\usepackage{indentfirst}
\newcommand\relphantom[1]{\mathrel{\phantom{#1}}}
\IEEEoverridecommandlockouts
\begin{document}

\title{Energy-Efficient Power Allocation for Secure Communications in Large-Scale MIMO Relaying Systems}
\author{\authorblockN{Jian~Chen, Xiaoming~Chen, Tao~Liu, and Lei~Lei
\\College of Electronic and Information Engineering, Nanjing
University of Aeronautics and Astronautics, China.
\\Email: $\{$chenjian04, chenxiaoming, tliu, leilei$\}$@nuaa.edu.cn
}} \maketitle

\begin{abstract}
In this paper, we address the problem of energy-efficient power
allocation for secure communications in an amplify-and-forward (AF)
large-scale multiple-input multiple-output (LS-MIMO) relaying system
in presence of a passive eavesdropper. The benefits of an AF LS-MIMO
relay are exploited to significantly improve the secrecy
performance, especially the secrecy energy efficiency (bit per
Joule). We first analyze the impact of transmit power at the relay
on the secrecy outage capacity, and prove that the secrecy outage
capacity is a concave function of transmit power under very
practical assumptions, i.e. no eavesdropper channel state
information (CSI) and imperfect legitimate CSI. Then, we propose an
energy-efficient power allocation scheme to maximize the secrecy
energy efficiency. Finally, simulation results validate the
advantage of the proposed energy-efficient scheme compared to the
capacity maximization scheme.
\end{abstract}

\section{Introduction}
The open nature of wireless medium is usually exploited to improve
the performance through multiuser transmission, but also results in
the information leakage to an unintended user. Traditionally, the
problem of wireless security is addressed at the upper layers of the
protocol stack by using sophisticated encryption techniques. Thanks
to the seminal work of Wyner \cite{Wyner}, it is found that secure
communication could be realized only by physical layer techniques,
namely physical layer security.

From an information-theoretic viewpoint, physical layer security is
in essence to maximize the performance difference between the
legitimate channel and the eavesdropper channel \cite{SC1}
\cite{SC2}. Thus, it makes sense to impair the interception signal
and to enhance the legitimate signal simultaneously. Motivated by
this, a variety of advanced physical layer techniques are introduced
to improve the secrecy performance. Wherein, MIMO relaying
techniques have received considerable attentions \cite{MIMORelay1}
\cite{MIMORelay3}. In \cite{AF} and \cite{DF}, the optimal
beamforming schemes at the relay for commonly used
amplify-and-forward (AF) and decode-and-forward (DF) relaying
protocols were presented in two-hop secure communications. Note that
the above schemes require global channel state information (CSI) to
design the transmit beams. In fact, the CSI, especially eavesdropper
CSI, is difficult to be obtained, since the eavesdropper is usually
passive and keeps silence. Then, a robust beamforming scheme was
given in \cite{RobustBeamforming}, assuming imperfect eavesdropper
CSI at the relay. Furthermore, if there is no any eavesdropper CSI,
a joint beamforming and jamming scheme was proposed in
\cite{Jamming}. Through transmitting artificial noise in the null
space of the legitimate channel, the interception signal is
weakened, while there is no effect on the legitimate signal.

Note that the joint beamforming and jamming scheme consumes extra
power to transmit artificial noise, resulting in a low energy
efficiency. Moreover, if legitimate CSI is imperfect, the artificial
noise will also affect the legitimate signal. Thus, it is necessary
to introduce new MIMO relaying techniques to further enhance
wireless security under very practical assumptions, i.e. no
eavesdropper CSI and imperfect legitimate CSI. Recently, it is found
that large-scale MIMO (LS-MIMO) can produce high-resolution spatial
beam, so as to avoid the information leakage to the unintended user
\cite{LS-MIMO}. In \cite{LS-MIMORelaying}, the secrecy performance
of LS-MIMO relaying techniques in secure communications was
analyzed. It was shown that even in very adverse environment, such
as short-distance interception, the secrecy performance can be
improved significantly. In secure communication, the transmit power
has a complicate effect on the secrecy performance, especially in
relaying systems. This is because the power will affect the
legitimate and the eavesdropper signals simultaneously. Thus, it
makes sense to choose an optimal power at the relay. Considering
that energy efficiency is a pivotal metric in wireless
communications, especially in secure communications
\cite{EnergyEfficiency} \cite{Huazi}, this paper focuses on
designing an energy-efficient power allocation scheme for secure
communication in an AF LS-MIMO relaying system. The contributions of
this paper are two-fold:

\begin{enumerate}

\item We derive the secrecy energy efficiency (bit per Joule) of
an AF LS-MIMO relaying system under imperfect CSI, and reveal the
impact of transmit power on the secrecy energy efficiency.

\item We propose an energy-efficient power allocation scheme by maximizing
the secrecy energy efficiency.

\end{enumerate}

The remainder of this paper is organized as follows. The two-hop
LS-MIMO relaying model and AF relaying protocol are introduced in
Section II. In Section III, we propose an energy-efficient power
allocation scheme. In Section IV, we present some simulation results
to validate the effectiveness of the proposed scheme. Finally, we
conclude the whole paper in Section V.

\section{System Model}
\begin{figure}[h] \centering
\includegraphics [width=0.4\textwidth] {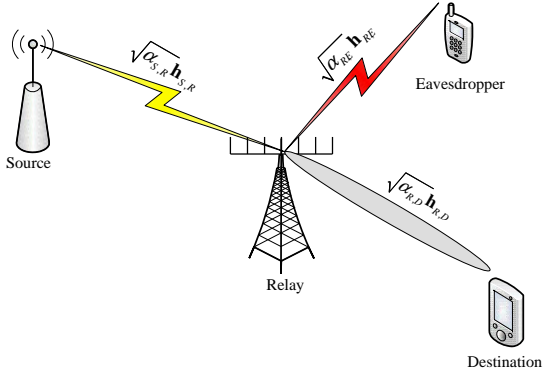}
\caption {The secure LS-MIMO relaying system.} \label{Fig1}
\end{figure}

In this section, we present the secure relaying system and the AF
relaying protocol under consideration.

\subsection{Secure Relaying System}
We consider a time division duplex (TDD) two-hop LS-MIMO relaying
system depicted in Fig.\ref{Fig1}, where a single antenna source
sends message to a single antenna destination with the aid of a
relay equipped with $N_R$ antennas, while a single antenna passive
eavesdropper intends to intercept the information. Note that $N_R$
is pretty large in such an LS-MIMO relaying system, i.e. $N_R=100$
or even larger. Due to a long propagation distance between the
source and the destination, we assume there is no direct
transmission between them. In other words, any successful
information transmission from the source to the destination must get
the help of the relay. It is assumed that the eavesdropper is far
away from the source and is close to the relay, since it thought the
signal comes from the relay. Note that it is a common assumption in
related literature \cite{SecureRelay}, since it is difficult for the
eavesdropper to overhear the signals from the source and the relay
simultaneously. Then, the eavesdropper only monitors the
transmission from the relay to the destination.

We use $\sqrt{\alpha_{i,j}}\textbf{h}_{i,j}$ to denote the channel
from node $i$ to $j$, where $i\in\{S,R\}$, $j\in\{R,D,E\}$ with $S$,
$R$, $D$ and $E$ representing the source, the relay, the destination
and the eavesdropper, respectively. $\alpha_{i,j}$ is the
distance-dependent path loss and $\textbf{h}_{i,j}$ is the
small-scale channel fading vector with independent and identically
distributed (i.i.d) zero mean and unit variance complex Gaussian
entries. It is assumed that $\alpha_{i,j}$ remains unchanged during
a relative long period and $\textbf{h}_{i,j}$ is block fading.

It takes two time slots to complete a whole transmission. In the
first time slot, the source transmits the information to the relay.
Thus, the received signal at the relay can be expressed as
\begin{equation}
\textbf{y}_R=\sqrt{P_S\alpha_{S,R}}\textbf{h}_{S,R}s+\textbf{n}_R,\label{eqn1}
\end{equation}
where $s$ is the normalized transmit signal, $P_S$ is the transmit
power at the source, $\textbf{n}_R$ is the additive Gaussian white
noise with unit variance at the relay. During the second time slot,
the relay forwards the post-processed signal $\textbf{r}$ to the
destination. Thus, the received signals at the destination and the
eavesdropper can be expressed as
\begin{equation}
y_D=\sqrt{P_R\alpha_{R,D}}\textbf{h}_{R,D}^H\textbf{r}+n_D,\label{eqn2}
\end{equation}
and
\begin{equation}
y_E=\sqrt{P_R\alpha_{R,E}}\textbf{h}_{R,E}^H\textbf{r}+n_{E},\label{eqn3}
\end{equation}
respectively, where $P_R$ is the transmit power at the relay, $n_D$
and $n_{E}$ are the additive Gaussian white noises with unit
variance at the destination and the eavesdropper, respectively.

\subsection{AF Relaying Protocol}
The relay adopts an AF relaying protocol to forward the information.
Then, the normalized signal to be transmitted at the relay is given
by
\begin{equation}
\textbf{r}=\textbf{F}\textbf{y}_{R},\label{eqn4}
\end{equation}
where $\textbf{F}$ is a transform matrix used at the relay.

We assume that the relay has perfect CSI $\textbf{h}_{S,R}$ by
channel estimation and gets partial CSI $\textbf{h}_{R,D}$ due to
channel reciprocity in TDD systems. The relation between the
estimated CSI $\hat{\textbf{h}}_{R,D}$ and the real CSI
$\textbf{h}_{R,D}$ is given by
\begin{equation}
\textbf{h}_{R,D}=\sqrt{\rho}\hat{\textbf{h}}_{R,D}+\sqrt{1-\rho}\textbf{e},\label{eqn5}
\end{equation}
where $\textbf{e}$ is the estimation error noise vector with i.i.d.
zero mean and unit variance complex Gaussian entries, and is
independent of $\hat{\textbf{h}}_{R,D}$. $\rho$, scaling from $0$ to
$1$, is the correlation coefficient between $\hat{\textbf{h}}_{R,D}$
and $\textbf{h}_{R,D}$. In addition, since the eavesdropper is
usually passive and keeps silence, the CSI $\textbf{h}_{R,E}$ is
unavailable at the relay. Therefore, $\textbf{F}$ is designed only
based on $\textbf{h}_{S,R}$ and $\hat{\textbf{h}}_{R,D}$, but is
independent of $\textbf{h}_{R,E}$. Considering the low complexity
and good performance in LS-MIMO systems, we combine maximum ratio
combination (MRC) and maximum ratio transmission (MRT) at the relay
to process the received signal. Thus, the transform matrix
$\textbf{F}$ is given by
\begin{equation}
\textbf{F}=\frac{\hat{\textbf{h}}_{R,D}}{\|\hat{\textbf{h}}_{R,D}\|}\frac{1}{\sqrt{P_S\alpha_{S,R}\|\textbf{h}_{S,R}\|^2+1}}\frac{\textbf{h}_{S,R}^H}{\|\textbf{h}_{S,R}\|}.\label{eqn6}
\end{equation}

In this paper, we adopt the secrecy outage capacity $C_{SOC}$ to
evaluate wireless security, since it is impossible to provide a
steady secrecy rate over fading channels if there is no eavesdropper
CSI at the relay. Secrecy outage capacity is defined as the maximum
available rate under the condition that the outage probability of
the real transmission rate being greater than secrecy capacity is
equal to a given value. Mathematically, it is given by
\begin{equation}
P_r(C_{soc}>C_D-C_E)=\varepsilon,\label{eqn7}
\end{equation}
where $C_D$ and $C_E$ are the legitimate and eavesdropper channel
capacities, respectively. $\varepsilon$ is an outage probability
associated to a secrecy outage capacity $C_{soc}$.


\section{Energy-Efficient Power Allocation}
In this section, we first present the secrecy outage capacity for an
AF LS-MIMO relaying system with imperfect CSI, analyze the impact of
transmit power at the relay on the secrecy outage capacity, and
finally derive a power allocation scheme by maximizing the secrecy
energy efficiency, namely a ratio of secrecy outage capacity over
total power consumption.

Based on the received signals in (\ref{eqn2}) and (\ref{eqn3}), the
legitimate and eavesdropper channel capacities are given by
\begin{equation}
C_D=W\log_2(1+\gamma_D),
\end{equation}
and
\begin{equation}
C_E=W\log_2(1+\gamma_E),
\end{equation}
respectively, where $W$ is the spectral bandwidth. $\gamma_D$ and
$\gamma_E$ are the signal-to-noise ratios (SNR) at the destination
and the eavesdropper, which can be expressed as
\begin{equation}
\gamma_D=\frac{P_SP_R\alpha_{S,R}\alpha_{R,D}|\textbf{h}_{R,D}^H\hat{\textbf{h}}_{R,D}|^2\|\textbf{h}_{S,R}\|^2}{P_R\alpha_{R,D}|\textbf{h}_{R,D}^H\hat{\textbf{h}}_{R,D}|^2+\|\hat{\textbf{h}}_{R,D}\|^2(P_S\alpha_{S,R}\|\textbf{h}_{S,R}\|^2+1)},
\end{equation}
and
\begin{equation}
\gamma_E=\frac{P_SP_R\alpha_{S,R}\alpha_{R,E}
|\textbf{h}_{R,E}^H\hat{\textbf{h}}_{R,D}|^2\|\textbf{h}_{S,R}\|^2}
{P_R\alpha_{R,E}|\textbf{h}_{R,E}^H\hat{\textbf{h}}_{R,D}|^2
+\|\hat{\textbf{h}}_{R,D}\|^2(P_S\alpha_{S,R}\|\textbf{h}_{S,R}\|^2+1)}.
\end{equation}

Thus, for the secrecy outage capacity, we have the following lemma:

\emph{Lemma 1}: For a given outage probability by $\varepsilon$, the
secrecy outage capacity of an AF LS-MIMO relaying system with
imperfect CSI can be expressed as
$C_{soc}(P_R)=W\log_2\left(1+\frac{P_SP_R\alpha_{S,R}\alpha_{R,D}\rho
N_R^2}{P_R\alpha_{R,D}\rho N_R+P_S\alpha_{S,R}N_R+1}\right)
-W\log_2\left(1+\frac{P_SP_R\alpha_{S,R}\alpha_{R,E}N_R\ln\varepsilon}{P_R\alpha_{R,E}\ln\varepsilon-P_S\alpha_{S,R}N_R-1}\right)$.

\begin{proof}
The secrecy outage capacity can be obtained based on (\ref{eqn7}) by
making use of the property of channel hardening in LS-MIMO systems
\cite{ChannelHardening}. We omit the proof due to space limitation,
and the detail can be referred to our previous work
\cite{LS-MIMORelaying}.
\end{proof}

Let $\rho\alpha_{R,D}N_R=A$, $-\alpha_{R,E}\ln\varepsilon=A\cdot
r_{l}$, $P_S\alpha_{S,R}N_R=B$, where
$r_{l}=\frac{-\alpha_{R,E}\ln\varepsilon}{\rho\alpha_{R,D}N_R}$ is
defined as the relative distance-dependent path loss. Then, the
secrecy outage capacity can be rewritten as
\begin{eqnarray}
C_{soc}(P_R)&=&W\log_2\left(1+\frac{P_RAB}{P_RA+B+1}\right)\nonumber\\
&&-W\log_2\left(1+\frac{P_RAB}{P_RA+\frac{B+1}{r_l}}\right).\label{eqn12}
\end{eqnarray}
Examining (\ref{eqn12}), it is found that if and only if $0<r_l<1$,
the secrecy outage capacity is positive. Obviously, only when
$C_{soc}$ is positive, the problem of energy efficiency makes sense.
In what follows, we only consider the case of $0<r_l<1$.

Prior to designing an energy-efficient power allocation scheme, we
first investigate the total power consumption in such a secure
relaying system. Herein, we model the total power $D$ as the sum of
powers at the source and the relay, which is given by
\begin{eqnarray}
D(P_R)&=&\frac{1}{2}P_S+\frac{1}{2}P_R+P_C,\label{eqn13}
\end{eqnarray}
where $P_C$ is the constant circuit power consumption at both the
relay and the source, including the power dissipations in the
transmitter filter, mixer, frequency synthesizer, digital-to-analog
converter and so on, which are independent of the actual transmit
signals. The factor 1/2 before $P_S$ and $P_R$ appears since the
source and the relay only send message in one slot, respectively. We
further assume that the transmit power $P_S$ at the source is fixed,
and focus on power allocation at the relay.

Hence, the secrecy energy efficiency for such a secure relaying
system is defined as the average number of bit per Joule securely
delivered to the destination. In this paper, we aim to maximize the
secrecy energy efficiency by distributing the transmit power at the
relay. Mathematically, the power allocation at the relay is
equivalent to the following optimization problem
\begin{align}
J_1:\quad &\max_{P_R}\quad{\frac{\frac{1}{2}C_{soc}(P_R)}{D(P_R)}}\label{eqn14}\\
& \begin{array}{r@{\quad}r@{}l@{\quad}l}
s.t.&P_R\leq P_T\label{eqn15}\\
\end{array} .
\end{align}
where $P_T$ is the transmit power constraint at the relay.
(\ref{eqn14}) is the so called secrecy energy efficiency.
(\ref{eqn15}) is the transmit power constraint at the relay. To
solve the above optimization problem, it is better to know the
monotonicity of $C_{soc}(P_R)$. For the secrecy outage capacity, we
have the following property:

\emph{Proposition 1}: The secrecy outage capacity of the LS-MIMO
relaying system increases when
$P_R\in\left(0,\sqrt{-\frac{P_S\alpha_{S,R}N_R+1}{\alpha_{R,E}\alpha_{R,D}\rho
N_R\ln\varepsilon}}\right)$, and decreases when
$P_R\in\left(\sqrt{-\frac{P_S\alpha_{S,R}N_R+1}{\alpha_{R,E}\alpha_{R,D}\rho
N_R\ln\varepsilon}},\infty\right)$.

\begin{proof}
Please refer to Appendix I.
\end{proof}

From Proposition 1, it is known that the maximum secrecy energy
efficiency must appear when
$P_R\in\left(0,\sqrt{-\frac{P_S\alpha_{S,R}N_R+1}{\alpha_{R,E}\alpha_{R,D}\rho
N_R\ln\varepsilon}}\right)$. This is because the secrecy outage
capacity is a decreasing function when $P_R$ belongs to
$\left(\sqrt{-\frac{P_S\alpha_{S,R}N_R+1}{\alpha_{R,E}\alpha_{R,D}\rho
N_R\ln\varepsilon}},\infty\right)$. Then, as $P_R$ increases, the
secrecy energy efficiency decreases. Hence, the optimization problem
$J_1$ can be rewritten as
\begin{align}
J_2:\quad&\max_{P_R}\quad{\frac{C_{soc}(P_R)}{P_S+P_R+2P_C}}\label{eqn16}\\
& \begin{array}{r@{\quad}r@{}l@{\quad}l}
s.t.&P_R\leq P_{\min}\label{eqn17}\\
\end{array} .
\end{align}
where
$P_{\min}=\min\left(P_T,\sqrt{-\frac{P_S\alpha_{S,R}N_R+1}{\alpha_{R,E}\alpha_{R,D}\rho
N_R\ln\varepsilon}}\right)$. The objective function (\ref{eqn16}) in
a nonlinear fractional manner is usually non-convex. In general, by
making use of the property of fractional programming
\cite{Dinkelbach}, the objective function is equivalent to
$C_{soc}(P_R)-q^*(P_S+P_R+2P_C)$, where $q^*$ is the maximum secrecy
energy efficiency, namely
$q^*=\max_{P_R}{\frac{C_{soc}(P_R)}{P_S+P_R+2P_C}}$. Thus, the
optimization problem $J_2$ is transformed as
\begin{align}
J_3:\quad&\max_{P_R}\quad{C_{soc}(P_R)-q^*(P_S+P_R+2P_C)}\label{eqn18}\\
& \begin{array}{r@{\quad}r@{}l@{\quad}l} s.t. &-P_R+P_{\min}\geq0.
\label{eqn19}
\end{array}
\end{align}
Checking the convexity of (\ref{eqn18}), we have the following
proposition:

\emph{Proposition 2}: $C_{soc}(P_R)-q^*(P_S+P_R+2P_C)$ is a concave
function of $P_R$, $\forall$
$P_R\in\left(0,\sqrt{-\frac{P_S\alpha_{S,R}N_R+1}{\alpha_{R,E}\alpha_{R,D}\rho
N_R\ln\varepsilon}}\right)$.

\begin{proof}
Please refer to Appendix II.
\end{proof}
According to Proposition 2, $J_3$ can be solved by the Lagrange
multiplier method. Firstly, the Lagrange dual function can be
written as
\begin{eqnarray}
\mathcal{L}(P_R,\theta)&=&C_{soc}(P_R)-q^*(P_S+P_R+2P_C)\nonumber\\
&&-\theta(P_R-P_{\min}),\label{eqn21}
\end{eqnarray}
where $\theta$ is the Lagrange multiplier corresponding to the
constraint (\ref{eqn19}). Therefore, the dual problem of $J_3$ is
given by
\begin{eqnarray}
\min_{\theta}{\max_{P_R}\quad{\mathcal{L}(P_R,\theta)}}.\label{eqn22}
\end{eqnarray}

For a given $\theta$, the optimal transmit power $P_R^*$ can be
derived by solving the following KKT condition
\begin{eqnarray}
\frac{\partial \mathcal{L}(P_R,\theta)}{\partial P_R}&=&\frac{\partial C_{soc}(P_R)}{\partial P_R}-q^*-\theta\nonumber\\
&=&0.\label{eqn23}
\end{eqnarray}
In addition, $\theta$ can be updated by the gradient method, which
is given by
\begin{eqnarray}
\theta(n+1)&=&[\theta(n)-\triangle_{\theta}(-P_R+P_{\min})]^+,\label{eqn24}
\end{eqnarray}
where $n$ is the iteration index, and $\triangle_{\theta}$ is the
positive iteration step. Above all, we propose an iteration
algorithm for energy-efficient power allocation at the relay as
follows: \rule{8.8cm}{1pt} Algorithm 1:Energy-Efficient Power
Allocation. \rule{8.8cm}{0.5pt}
\begin{enumerate}
\item Initialize the maximum number of iterations $L_{\max}$ and
the maximum tolerance $\delta$.

\item  Set a maximum energy efficiency $q=0$ and iteration index $n=0$.

\item  Figure out the solution $P'$  of (\ref{eqn23}) for a given
$q$.

\item  Update $\theta$  according to (\ref{eqn24}) and let $n=n+1$.

\item  If $C_{soc}(P_R)-q(P_S+P'+2P_C)<\delta$, then return
$P_R^*=P'$ and $q^*=\frac{C_{soc}(P_R)}{P_S+P'+2P_C}$. Otherwise, if
$n<L_{\max}$, go to 3) with $q=\frac{C_{soc}(P_R)}{P_S+P'+2P_C}$.
\end{enumerate}
\rule{8.8cm}{1pt}

\section{Simulation Results}
To examine the effectiveness of the proposed energy-efficient power
allocation scheme for an AF LS-MIMO relaying system, we present
several simulation results in the following scenarios: we set
$N_R=100$, $W=10$ KHz, $\rho=0.9$, $P_C=5$ dB, $P_T=10$ dB and
$\varepsilon=0.05$. We assume that the relay is in the middle of the
source and the destination. For the sake of calculational
simplicity, we normalize the path loss as
$\alpha_{S,R}=\alpha_{R,D}=1$ and use $\alpha_{S,E}$ to denote the
relative path loss. Note that $\alpha_{R,E}>1$ means the
eavesdropper is closer to the relay than the destination. We set
$\alpha_{R,E}=1.5$ without extra explanation. In the following
results, ``number of iterations" refers to the number of iterations
in Algorithm 1.

\begin{figure}[h] \centering
\includegraphics [width=0.5\textwidth] {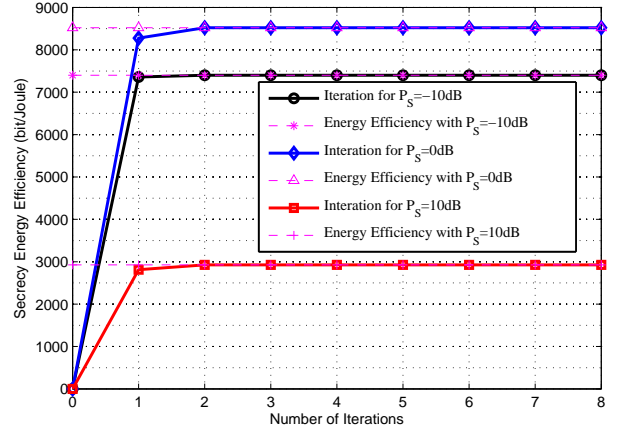}
\caption {Convergence speed with different transmit powers at the
source.} \label{Fig2}
\end{figure}

First, we illustrate the convergence speed of the proposed iterative
algorithm with different transmit powers at the source $P_S$. As
observed in Fig.\ref{Fig2}, the iterative algorithm converges to the
maximum energy efficiency within no more than 8 iterations, so the
algorithm is reliable and efficient. Moreover, it is found that the
energy efficiency is not a monotonically increasing function of
source transmit power $P_S$, since the maximum energy efficiency
with $P_S=0$ dB is bigger than that with $P_S=10$ dB. Then, it makes
sense to choose an optimal source transmit power. We will analyze
the optimal source transmit power in future work.

\begin{figure}[h] \centering
\includegraphics [width=0.5\textwidth] {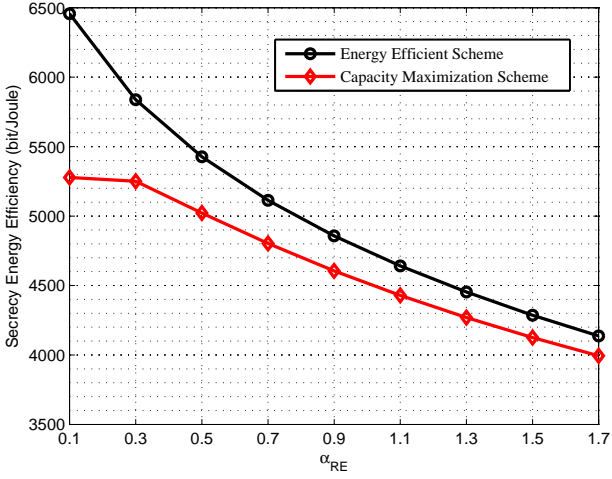}
\caption {Performance comparison of energy efficient and capacity
maximization power allocation schemes.} \label{Fig3}
\end{figure}

Then, we show the energy efficiency gain of the proposed
energy-efficient power allocation scheme compared to the secrecy
outage capacity maximization scheme with $P_S=10$ dB. As seen in
Fig\ref{Fig3}, the proposed energy-efficient scheme obviously
performs better than the capacity maximization scheme, especially at
small $\alpha_{R,E}$ regime. For example, at $\alpha_{R,E}=0.1$,
there is about 1.2 Kb/J gain. As $\alpha_{R,E}$ increases, the gain
becomes smaller. The reasons are two-fold. On the one hand, as
$\alpha_{R,E}$ increases, the secrecy outage capacity decreases
accordingly. On the other hand, the feasible set of $P_R$ is also
reduced.

\begin{figure}[h] \centering
\includegraphics [width=0.5\textwidth] {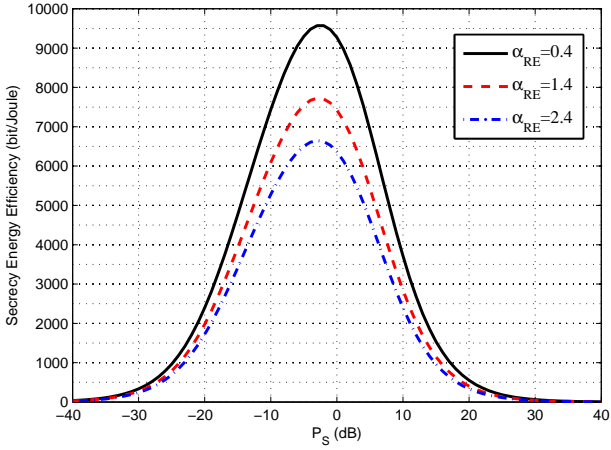}
\caption {Maximum energy efficiency with different $P_S$.} \label{Fig4}
\end{figure}

Next, we examine the effect of $P_S$ on the maximum energy
efficiency. As seen in Fig.\ref{Fig4}, the maximum energy efficiency
is a concave function of $P_S$. As $P_S$ tends to zero, the maximum
energy efficiencies with different $\alpha_{R,E}$ asymptotically
approach zero. This is because the secrecy outage capacity is
approximately zero and the total power assumption is nonzero under
such a condition. At high $P_S$ region, the maximum energy
efficiency also approaches zero, since the secrecy outage capacity
will saturated if $P_S$ is sufficiently large and the total power
consumption is quite large. Hence, there is an optimal source
transmit power in the sense of maximizing the secrecy energy
efficiency.

\begin{figure}[h] \centering
\includegraphics [width=0.5\textwidth] {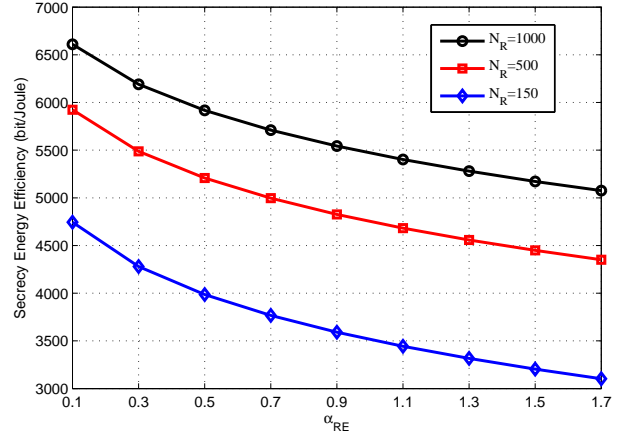}
\caption {Performance of the proposed energy-efficient power
allocation scheme with different numbers of antennas.} \label{Fig5}
\end{figure}

Finally, we investigate the function of the number of antennas at
the relay on the energy efficiency of the proposed scheme with
$P_S=10$ dB. As shown in Fig.\ref{Fig5}, with the increase of
$\alpha_{R,E}$, all the maximum energy efficiencies with different
numbers of relay antennas decrease. However, for a given
$\alpha_{R,E}$, as $N_R$ adds, the maximum energy efficiency
increases significantly. Hence, we can increase the secrecy energy
efficiency by simply adding the antennas at the relay, which is a
main advantage of LS-MIMO relaying systems.

\section{Conclusion}
This paper proposed an energy-efficient power allocation scheme for
an AF LS-MIMO relaying system with imperfect CSI. The
energy-efficiency scheme has a fast convergence characteristics and
obviously outperforms better than the capacity maximization scheme.
More importantly, it is feasible to obtain a high secrecy energy
efficiency even in the case of short-distance interception by adding
the number of antennas at the relay.

\begin{appendices}
\section{Proof of Proposition 1}
At first, we take derivative of (\ref{eqn12}) with respect to $P_R$,
which is given by (\ref{equ19}) at the top of the next page.
\begin{figure*}
\begin{equation}
C_{soc}^{'}=\frac{W}{\ln2}B(1+B)\left(\frac{A}{(P_RA+B+1)^2+P_RAB(P_RA+B+1)}-\frac{Ar_{l}}{(P_RAr_{l}+B+1)^2+P_RABr_{l}(P_RAr_{l}+B+1)}\right).\label{equ19}
\end{equation}
\end{figure*}
Let $C_{soc}^{'}(P_R)=0$, we get two solutions
\begin{eqnarray}
P_R&=&\frac{1}{Ar_{l}}\sqrt{r_{l}(B+1)},\label{eqn20}
\end{eqnarray}
and
\begin{eqnarray}
P_R&=&-\frac{1}{Ar_{l}}\sqrt{r_{l}(B+1)}.\label{eqn21}
\end{eqnarray}

Considering $P_R>0$, (\ref{eqn20}) is the unique optimal solution in
this case. What's more, when
$P_R<\frac{1}{Ar_{l}}\sqrt{r_{l}(B+1)}$, we have $C_{soc}'(P_R)>0$.
Otherwise, if $P_R>\frac{1}{Ar_{l}}\sqrt{r_{l}(B+1)}$, we have
$C_{soc}'(P_R)<0$. Specifically, $C_{soc}(P_R)$ improves as $P_R$ increases in
the region from $0$ to $\frac{1}{Ar_{l}}\sqrt{r_{l}(B+1)}$, while
$C_{soc}(P_R)$ decreases as $P_R$ increases in the region from
$\frac{1}{Ar_{l}}\sqrt{r_{l}(B+1)}$ to infinity. Hence, we get the Proposition 1.

\section{Proof of Proposition 2}
Obviously, $C_{soc}(P_R)-q^*(P_S+P_R+2P_C)$ has the same convexity
or concavity as $C_{soc}(P_R)$, since $q^*(P_S+P_R+2P_C)$ is affine.
The proof that $C_{soc}(P_R)$ is concave is given as follows:
\begin{eqnarray}
C_{soc}(P_R)&=&W\log_2\left(1+\frac{P_RAB}{P_RA+B+1}\right)\nonumber\\
&&-W\log_2\left(1+\frac{P_RABr_l}{P_RAr_l+B+1}\right)\nonumber\\
&=&W\log_2\left(\frac{1+\frac{P_RAB}{P_RA+B+1}}{1+\frac{P_RABr_l}{P_RAr_l+B+1}}\right).
\end{eqnarray}

Let
$g_0(P_R)=\frac{1+\frac{P_RAB}{P_RA+B+1}}{1+\frac{P_RABr_l}{P_RAr_l+B+1}}$,
then we have
\begin{equation}
\begin{split}
g_0(P_R)&=\frac{1+\frac{P_RAB}{P_RA+B+1}}{1+\frac{P_RABr_l}{P_RAr_l+B+1}}\nonumber\\
&=\frac{P_RAB+P_RA+B+1}{P_RA+B+1}\\
&\relphantom{=}{}\times\frac{P_RAr_l+B+1}{P_RABr_L+P_RAr_l+B+1}\nonumber\\
&=\frac{(P_RA+1)(B+1)}{P_RA+B+1}\frac{P_RAr_l+B+1}{(P_RAr_l+1)(B+1)}\nonumber\\
&=\frac{P_RA+1}{P_RA+B+1}\frac{P_RAr_l+B+1}{P_RAr_l+1}\nonumber\\
&=1+\frac{B}{P_RAr_l+1}-\frac{B}{P_RA+B+1}\nonumber\\
&\relphantom{=}{}-\frac{B^2}{(P_RAr_l+1)(P_RAr_l+B+1)}.
\end{split}
\end{equation}

Let
$g_1(P_R)=\frac{1}{P_RAr_l+1}-\frac{1}{P_RA+B+1}-\frac{B}{(P_RAr_l+1)(P_RAr_l+B+1)}$.
Apparently, $g_1(P_R)$ has the same convexity or concavity as
$g_0(P_R)$ due to $B>0$.

For the sake of notational and computational simplicity, let
$P_1=P_RA$, then $g_1(P_R)$ can be written as
$g_2(P_1)=\frac{1}{P_1r_l+1}-\frac{1}{P_1+B+1}-\frac{B}{(P_1r_l+1)(P_1+B+1)}$.
Because of $A>0$, $g_2(P_1)$ preserves the convexity or concavity of
$g_1(P_R)$, which has the same convexity or concavity as $g_0(P_R)$.
In other words, $g_2(P_1)$ and $g_0(P_R)$ have the same convexity or
concavity. By some simplification, $g_2(P_1)$ can be rewritten as
\begin{eqnarray}
g_2(P_1)&=&\frac{(1-r_l)P_1}{(P_1r_l+1)(P_1+B+1)}.
\end{eqnarray}


Now, we prove the convexity of $g_2(P_1)$. Firstly, we take second
derivative of $g_2(P_1)$ with respect to $P_1$, which is given by
\begin{equation}
\begin{split}
g_2''(P_1)&=\frac{P_1r_l(P_1^2r_l-3B-3)-B^2r_l-2Br_l-r_l-B-1}{\left((P_1r_l+1)(P_1+B+1)\right)^3}\\
&\relphantom{=}{}\times(1-r_l).\label{eqn27}
\end{split}
\end{equation}
Because of $B>0$, $P_1>0$ and $0<r_l<1$, $-B^2r_l-2Br_l-r_l-B-1$
must be negative, and
$\frac{1-r_l}{\left((P_1r_l+1)(P_1+B+1)\right)^3}$ must be positive.
Then, once $P_1\in\left(0,\sqrt{\frac{3(B+1)}{r_l}}\right)$,
$g_2''(P_1)$ is negative. In other words, $g_2(P_1)$ is concave when
$P_1\in\left(0,\sqrt{\frac{3(B+1)}{r_l}}\right)$. Therefore,
$g_0(P_R)$ is also concave when
$P_R\in\left(0,\frac{1}{A}\sqrt{\frac{3(B+1)}{r_l}}\right)$. Thus,
$W\log_2\left(g_0(P_R)\right)$, namely $C_{soc}(P_R)$, is a concave
function due to
$C_{soc}''(P_R)=-W\frac{1}{g_0(P_R)^2}g_0'(P_R)^2+W\frac{1}{g_0(P_R)}g_0''(P_R)\leq0$.
Above all, $C_{soc}(P_R)-q^*(P_S+P_R+2P_C)$ is concave. Hence, we
get the Proposition 2.

\end{appendices}

\end{document}